\documentstyle[prb,aps,12pt]{revtex}
\begin{document}
\title{Polaron features for long-range electron-phonon interaction}
\author{C. A. Perroni, V. Cataudella, and G. De Filippis}
\address{Coherentia-INFM  and Dipartimento di Scienze Fisiche, \\
Universit\`{a} degli Studi di Napoli ``Federico II'',\\
Complesso Universitario Monte Sant'Angelo,\\
Via Cintia, I-80126 Napoli, Italy}
\maketitle

\begin {abstract}

The polaron features for long-range electron-phonon interaction
are investigated by extending a variational approach previously
proposed for the study of systems with local coupling. The
ground-state spectral weight, the average kinetic energy, the mean
number of phonons, and the electron-lattice correlation function
are discussed for a wide range of model parameters focusing on the
adiabatic regime and comparing the results with the short-range
case (Holstein model). A strong mixing of electronic and phononic
degrees of freedom for small values of the electron-phonon
coupling constant is found in the adiabatic case due to the
long-range interaction. Finally a polaron "phase diagram" is
proposed.

\end {abstract}


\newpage

\section {Introduction}

In the last years many experimental results have pointed out the
presence of strong electron-phonon ($el-ph$) coupling and
polaronic effects in several compounds, such as high-temperature
cuprate superconductors and colossal magnetoresistance manganites.
\cite{1,2} This large amount of experimental data has renewed the
interest in studying simplified $el-ph$ coupled systems of the
Holstein \cite{4} or Fr$\ddot{o}$hlich \cite{froh} type and in
proposing more realistic interaction models. \cite{alex,sandvik}

The Holstein molecular crystal model is the prototype for
short-range ($SR$) $el-ph$ interaction since it takes into account
local coupling of a tight-binding electron to optical local phonon
modes. Till now an exact solution for this model has not been
found and perturbative expansions are not able to describe the
most interesting region characterized by intermediate $el-ph$
couplings and electron and phonon energy scales not well
separated. In this regime, as shown by several numerical studies
\cite{7,korn,8,10,ciuchi} and variational approaches,
\cite{11,17,17bis} the system undergoes a crossover from a weakly
dressed electron to a massive localized polaronic quasiparticle,
the small Holstein polaron ($SHP$), with increasing the strength
of interaction. All the ground state properties of the Holstein
moldel can be described with great accuracy by a variational
approach \cite{17,17bis} based on a linear superposition of Bloch
states that describe weak and strong coupling polaron wave
functions. Indeed this method provides an immediate physical
interpretation of the intermediate regime and is in excellent
agreement with numerical results.

Recently a quite general $el-ph$ lattice Hamiltonian with a
"density-displacement" type interaction has been introduced in
order to understand the role of long-range ($LR$) coupling on the
polaron formation. \cite{alex,alex1} The model for a single
particle is described by the Hamiltonian $H$

\begin{eqnarray}
H=-t\sum_{<i,j>} c^{\dagger}_{i}c_{j} +\omega_0\sum_{i} \left(
a^{\dagger}_i a_i +\frac{1}{2} \right) +\alpha \omega_0 \sum_{i,j}
f(|\vec{R}_i-\vec{R}_j|) c^{\dagger}_{i}c_{i}\left(
a_j+a^{\dagger}_j\right). \label{1r}
\end{eqnarray}
The units are such that $\hbar=1$. In Eq.(\ref{1r})
$c^{\dagger}_{i}$ ($c_i$) denotes the electron creation
(annihilation) operator at site $i$, whose position vector is
indicated by $\vec{R}_{i}$, and the symbol $<>$ denotes nearest
neighbours linked through the transfer integral $t$. The operator
$a^{\dagger}_i$ ($a_i$) represents the creation (annihilation)
operator for phonon on the site $i$, $\omega_0$ is the frequency
of the optical local phonon modes, $\alpha$ controls the strength
of $el-ph$ coupling, and $f(|\vec{R}_i-\vec{R}_j| )$ is the
interacting force between an electron on the site $i$ and an ion
displacement on the site $j$.

The Hamiltonian (\ref{1r}) reduces to the Holstein model if
$f(|\vec{R}_i-\vec{R}_j| )=\delta_{\vec{R}_i,\vec{R}_j}$, while in
general it contains $LR$ interaction. In particular when one
attempts to mimic the nonscreened coupling between doped holes and
apical oxygen in some cuprates, \cite {alex} the interaction force
is given by
\begin{equation}
f(|\vec{R}_i-\vec{R}_j|)=  \left(|\vec{R}_i-\vec{R}_j|^{2} +1
\right)^{-\frac{3}{2}}, \label{force}
\end {equation}
if the distance $|\vec{R}_i-\vec{R}_j|$ is measured in units of
lattice constant. Considering the general $el-ph$ matrix element
$M_{\vec{q}}$
\begin{equation}
M_{\vec{q}}=\frac{\alpha \omega_0}{\sqrt{L}} \sum_{m}
f(|\vec{R}_m|) e^{i \vec{q} \cdot \vec{R}_m},
\end{equation}
with $L$ number of lattice sites, we can define the polaronic
shift $E_p$
\begin{equation}
E_p=\sum_{\vec{q}}\frac{M^{2}_{\vec{q}}}{\omega_0},
\end{equation}
and the coupling constant $\lambda=E_p/zt$, with $z$ lattice
coordination number, that represents a natural measure of the
strength of the $el-ph$ interaction in both $SR$ and $LR$ case.
Clearly for $LR$ interaction forces the matrix element
$M_{\vec{q}}$ is peaked around $\vec{q}=0$. Since it has been
claimed that the enhancement of the forward direction in the
$el-ph$ scattering could play a role in explaining several
anomalous properties of cuprates as the linear temperature
behaviour of the resistivity and the {\it d}-wave symmetry of the
superconducting gap, \cite{cappel,kulic} the study of lattice
polaron features for $LR$ interactions is important in order to
clarify the role of the $el-ph$ coupling in complex systems.

When the interaction force is given by Eq. (\ref{force}), the
model has been investigated applying a path-integral Monte-Carlo
($PIMC$) algorithm \cite{alex,alex1} efficient in the
thermodynamic limit. The first investigations have been mainly
limited to the determination of the polaron effective mass
pointing out that, due to the $LR$ coupling, the polaron is much
lighter than the $SHP$ with the same binding energy in the strong
coupling regime. Furthermore it has been found that this effect
due to the weaker band renormalization becomes smaller in the
antiadiabatic regime. Then the quasi-particle properties have been
studied by an exact Lanczos diagonalization method \cite{fehs} on
finite one-dimensional lattices (up to 10 sites) making a close
comparison with the corresponding properties of $HP$. As a result
of the $LR$ interaction, the lattice deformation induced by the
electron is spread over many lattice sites in the strong coupling
region giving rise to the formation of a large polaron ($LP$) as
in the weak coupling regime. All numerical and analytical results
have been mainly obtained in the antiadiabatic and non-adiabatic
regime. Only recently the behavior of the effective mass of a
two-site system \cite{alex2} in the adiabatic regime has been
studied within the nearest-neighbor approximation for the $el-ph$
interaction confirming that the $LP$ is lighter than the $SHP$ at
strong coupling.

In this paper we pursue the study of the ground state of the model
with the interaction force given by Eq. (\ref{force}) in the
thermodynamic limit. We employ a variational approach previously
proposed for the study of systems with $el-ph$ local coupling
\cite{17,17bis} and based on a linear superposition of Bloch
states that describe weak and strong coupling polaron wave
functions. Although the method is valid for any spatial dimension,
we limit our study to the one-dimensional case. It has been found
that the variational approach provides an estimate of the ground
state energy in good agreement with $PIMC$ results. The evolution
of the ground-state spectral weight, the average kinetic energy,
the mean number of phonons, and the electron-lattice correlation
function with respect to the adiabaticity ratio $\omega_0/t$ and
the $el-ph$ coupling constant is discussed focusing on the
adiabatic regime. Indeed, in the adiabatic case, there is a range
of values of the $el-ph$ coupling where the ground state is well
described by a particle with a weakly renormalized mass but a
spectral weight much smaller than unity. Furthermore, with
increasing the strength of interaction in the same regime, the
renormalized mass gradually increases, while the average kinetic
energy is not strongly reduced. Finally regions of the model
parameters are distinguished according to the values assumed by
the spectral weight. The resulting "phase diagram" \cite{phase}
shows strong mixing of electronic and phononic degrees of freedom
for small values of the $el-ph$ coupling constant in the adiabatic
case.

\section {Variational wave function}
The variational approach is summarized following the lines of
previous works. \cite{17,17bis}

We consider as trial wave functions translational invariant Bloch
states obtained by taking a superposition of localized states
centered on different lattice sites:
\begin{equation}
|\psi^{(i)}_{\vec{k}}>=\frac{1}{\sqrt{L}}\sum_{\vec{R}_n}e^{i\vec{k}\cdot
\vec{R}_n}|\psi^{(i)}_{\vec{k}}(\vec{R}_n)>,
 \label{12rn}
\end{equation}
where
\begin{equation}
|\psi^{(i)}_{\vec{k}}(\vec{R}_n)> = e^{\sum_{\vec{q}}\left[
h^{(i)}_{\vec{q}}(\vec{k})a_{\vec{q}} e^{i\vec{q}\cdot \vec{R}_n}
+h.c.\right]} \sum_m \phi^{(i)}_{\vec{k}}(\vec{R}_m)
c^{\dagger}_{m+n}|0> .
 \label{13rn}
\end{equation}
In Eq. (\ref{12rn}) the apex $i=w,s$ indicates the weak and strong
coupling polaron wave function, respectively, $|0>$ denotes the
electron and phonon vacuum state, and
$\phi^{(i)}_{\vec{k}}(\vec{R}_m)$ are variational parameters
defining the spatial broadening of the electronic wave function.
The phonon distribution functions $h^{(i)}_{\vec{q}}(\vec{k})$ are
chosen in order to reproduce the description of polaron features
in the two asymptotic limits. \cite{17} Therefore the weak
coupling phonon distribution function $h^{(w)}_{\vec{q}}(\vec{k})$
is assumed as
\begin{equation}
h^{(w)}_{\vec{q}}(\vec{k})=
\frac{M_{\vec{q}}}{\omega_0+E_b(\vec{k}+\vec{q})-E_b(\vec{k})},
\label{22rn}
\end{equation}
where $E_b(\vec{k})$ is the free electron band energy, while the
strong coupling phonon distribution function
$h^{(s)}_{\vec{q}}(\vec{k})$ as
\begin{equation}
h^{(s)}_{\vec{q}}(\vec{k})=
\frac{M_{\vec{q}}}{\omega_0}\sum_{m}|\phi_{\vec{k}}(\vec{R}_m)|^2
e^{i\vec{q}\cdot \vec{R}_m}.
\label{14rn}
\end{equation}

A careful inspection of weak and strong coupling wave functions
shows that in the intermediate regime they are not orthogonal and
the off-diagonal matrix elements of the Hamiltonian are not zero.
Hence the ground state energy is determined by considering as
trial state a linear superposition of the weak and strong coupling
wave functions:
\begin{equation}
|\psi_{\vec{k}}>=\frac{A_{\vec{k}}
|\overline{\psi}^{(w)}_{\vec{k}}>+ B_{\vec{k}}
|\overline{\psi}^{(s)}_{\vec{k}}>}
{\sqrt{A^2_{\vec{k}}+B^2_{\vec{k}}
+2A_{\vec{k}}B_{\vec{k}}S_{\vec{k}}}},
\label{31r}
\end{equation}
where
\begin{eqnarray}
&&|\overline{\psi}^{(w)}_{\vec{k}}>= \frac{|\psi^{(w)}_{\vec{k}}>}
{\sqrt{<\psi^{(w)}_{\vec{k}}|\psi^{(w)}_{\vec{k}}>}},
|\overline{\psi}^{(s)}_{\vec{k}}>= \frac{|\psi^{(s)}_{\vec{k}}>}
{\sqrt{<\psi^{(s)}_{\vec{k}}|\psi^{(s)}_{\vec{k}}>}} \label{32r}
\end{eqnarray}
and $S_{\vec{k}}$
\begin{equation}
S_{\vec{k}}=
\frac{<\overline{\psi}^{(w)}_{\vec{k}}|\overline{\psi}^{(s)}_{\vec{k}}>+h.c.}
{2}
\label{33r}
\end{equation}
is the overlap factor of the two wave functions
$|\overline{\psi}^{(w)}_{\vec{k}}>$ and
$|\overline{\psi}^{(s)}_{\vec{k}}>$. In Eq.(\ref{31r})
$A_{\vec{k}}$ and $B_{\vec{k}}$ are two additional variational
parameters which provide the relative weight of the weak and
strong coupling solutions of the system for any particular value
of $\vec{k}$.

We perform the minimization procedure with respect to the
parameters $\phi^{(w)}_{\vec{k}}(\vec{R}_m)$,
$\phi^{(s)}_{\vec{k}}(\vec{R}_m)$, $A_{\vec{k}}$ and
$B_{\vec{k}}$, assuming
\begin{equation}
\phi^{(i)}_{\vec{k}}(\vec{R}_n)=\alpha^{(i)}_{\vec{k}}
\delta_{\vec{R}_n,0}+ \beta^{(i)}_{\vec{k}} \sum_{\delta}
\delta_{\vec{R}_n,\vec{\delta}}+
\gamma^{(i)}_{\vec{k}}\sum_{\delta'
}\delta_{\vec{R}_n,\vec{\delta'}}+
\eta^{(i)}_{\vec{k}}\sum_{\delta''
}\delta_{\vec{R}_n,\vec{\delta''}}, \label{170r}
\end{equation}
where the quantities $\alpha^{(i)}_{\vec{k}}$,
$\beta^{(i)}_{\vec{k}}$, $\gamma^{(i)}_{\vec{k}}$, and
$\eta^{(i)}_{\vec{k}}$ denote variational parameters, and the
symbols $\delta$, $\delta'$, $\delta''$ indicate, respectively,
the nearest, the next-nearest neighbors and so on. This choice
takes into account the broadening of the electron wave functions
up to third neighbors and provides an accurate description of the
polaron features for any value of the parameters of the
Hamiltonian. The ground state energies obtained with this choice
are slightly higher than $PIMC$ mean energies, being the
difference less than $0.5 \%$ in the worst case of intermediate
regime. We note that these wave functions can be improved adding
further terms in Eq. (\ref{170r}), so it is possible to obtain
better and better estimates of the energy.

\section {Results}

In this paper we study the properties of the ground-state in the
one-dimensional case.

In Fig. 1(a) we report the polaron ground state energy for
different values of the adiabaticity ratio as a function of the
$el-ph$ constant coupling $\alpha$. We have checked that our
variational proposal recovers the asymptotic perturbative results
and improves significantly these asymptotic estimates in the
intermediate region. Moreover, our data for the ground-state
energy in the intermediate region are in very good agreement with
the results of the $PIMC$ approach \cite{alex} shown as diamonds
in Fig. 1(a). The consistency of the results with a numerically
more sophisticated approach indicates that the true wave function
is very close to a superposition of weak and strong coupling
states.

Another property of interest is the ground state spectral weight $Z$
\begin{equation}
Z=Z_{k=0}=|<\psi_{k=0}|c^{\dagger}_{k=0}|0>|^2, \label{6n}
\end{equation}
that gives the fraction of the bare electron state in the
polaronic trial wave function. It measures how much the
quasiparticle is different from the free electron ($Z=1$), and a
small value of $Z$ indicates a strong mixing of electronic and
phononic degrees of freedom. As plotted in Fig. 1(b), the increase
of the $el-ph$ coupling strength induces a decrease of the
spectral weight that is smooth also in the adiabatic regime. The
reduction of $Z$ is closely related to the decrease of the Drude
weight obtained by exact diagonalizations \cite{fehs} pointing out
a gradual suppression of coherent motion. We note that the
behavior of $Z$ is different from that of the local Holstein
model. In fact for the latter $Z$  results to be very close to the
ratio $m/m^*$, with $m$ and $m^*$ bare electron and effective
polaron mass, respectively, \cite{fehs} while for $LR$ couplings
$Z < m/m^*$ in the intermediate to strong coupling adiabatic
regime. This relation is confirmed by the results shown in Fig.1
(b), where the dash-dotted line and the squares on a similar line
indicate the spectral weight $Z$ and the ratio $m/m^*$ obtained
within the variational approach, respectively, as a function of
the coupling constant $\alpha$ at $\omega_0/t=0.25$. \cite{alex1}
Actually there is a large region of the parameters in the
adiabatic regime where the ground state is well described by a
particle with a weakly renormalized mass but a spectral weight $Z$
much smaller than unity. While the electron drags the phonon cloud
coherently through the lattice, with increasing the $el-ph$
coupling in the adiabatic case, a band collapse occurs in the $SR$
case, while the particle undergoes a weaker band renormalization
in the case of $LR$ interactions. Therefore in the $LR$ case the
polaron results lighter than the $SHP$ in the intermediate to
strong coupling adiabatic regime.

Insight about the electron state is obtained by calculating its
kinetic energy $K$ in units of the bare one. Since the average
kinetic energy gives the total weight of the optical conductivity,
$K$ includes both coherent and incoherent transport processes.
\cite{fehs} As reported in Fig. 1(b) and Fig. 1(c), in the strong
coupling adiabatic region before the electron is selftrapped ($K
\ll 1$), the average kinetic energy is weakly renormalized, the
ratio $m/m^*$ is reduced and the spectral weight is nearly zero.
Furthermore, in Figs. 1(c) and 1(d), within the adiabatic regime,
the average kinetic energy and the mean number of phonons do not
show any sharp change by increasing the $el-ph$ coupling.

Another quantity associated to the polaron formation is the
correlation function $S(R_l)$
\begin{equation}
S(R_l)=S_{k=0}(R_l)= \frac{\sum_{n} <\psi_{k=0}
|c^{\dagger}_nc_n\left(a^{\dagger}_{n+l}+a_{n+l}\right)|\psi_{k=0}>}
{<\psi_{k=0} |\psi_{k=0}>} \label{102r}
\end{equation}
or equivalently the normalized correlation function
$\chi(R_l)=S(R_l)/N$, with $N=\sum_l S(R_l)$. In Fig. 2(a) we
report the correlation function $S(R_l)$ at $\omega_0/t=1$ for
several values of the $el-ph$ interaction. The lattice deformation
is spread over many lattice sites giving rise to the formation of
$LP$ also in the strong coupling regime where really the
correlation function assumes the largest values. In the inset of
Fig. 2(a) the normalized electron-lattice correlation function
$\chi$ shows consistency with the corresponding quantity
calculated in a previous work. \cite{fehs} While in the weak
coupling regime the amplitude $\chi$ is smaller than the quantum
lattice fluctuations, increasing the strength of the interaction,
it becomes stronger and the lattice deformation is able to
generate an attractive potential that can trap the charge carrier.
Clearly, even if the correlations between electron and lattice are
large, the resulting polaron is delocalized over the lattice due
to the translational invariance. Finally the variation of the
lattice deformation as a function of $\omega_0/t$ shown in Fig.
2(b) can be understood as a retardation effect. In fact, for small
$\omega_0/t$, the less numerous phonons excited by the passage of
the electron take a long time to relax, therefore the lattice
deformation increases far away from the current position of the
electron.

In Fig. 3 we propose a "phase diagram" based on the values assumed
by the spectral weight in analogy with the Holstein polaron.
\cite{17bis} Analyzing the behavior of $Z$ it is possible to
distinguish three different regimes: (1) quasi-free-electron
regime ($0.9 < Z < 1$) where the electron has a weakly
renormalized mass and the motion is coherent; (2) crossover regime
($0.1 < Z < 0.9$) characterized by intermediate values of spectral
weight and a mass not strongly enhanced; (3) strong coupling
regime ($Z<0.1$) where the spectral weight is negligible and the
mass is large but not enormous. We note that for $LR$ interactions
in the adiabatic case there is strong mixing of electronic and
phononic degrees of freedom for values of the coupling constant
$\lambda$ (solid lines) smaller than those characteristic of local
Holstein interaction (dashed lines). Furthermore in this case,
entering the strong coupling regime, the charge carrier does not
undergo any abrupt localization, on the contrary, as indicated
also by the behavior of the average kinetic energy $K$, it is
quite mobile.

In order to study the effects of different $el-ph$ interactions,
we have evaluated the average kinetic energy for both $LR$ and
$SR$ cases. As reported in Fig. 4(a), in excellent agreement with
a previous study, \cite{fehs} for $LR$ interactions $K$ decreases
very gradually with increasing $\lambda$. Furthermore, if the
regime of parameters where the spectral weight $Z=0.1$ is
considered, as shown in Fig. 4(b), in the adiabatic case the
average kinetic energy is larger for $LR$ interactions (solid
line) with respect to local Holstein ones (dashed line). The
comparison emphasizes that due to $LR$ interactions in the
adiabatic regime $K$ is slightly renormalized even if the coherent
motion is small.

\section{Conclusions}

In this paper we have extended a previous variational approach in
order to study the polaronic ground-state features of a one
dimensional $el-ph$ model with long-range interaction. The trial
function is based on a linear superposition of Bloch states that
describe weak and strong coupling polaron wave functions and it
provides an estimate of the ground state energy in good agreement
with numerical methods. The results relative to spectral weight,
the average kinetic energy, the mean number of phonons, and the
electron-lattice correlation function have been discussed mainly
in the adiabatic regime. It has been possible to identify a range
of intermediate values of the $el-ph$ coupling constant in the
adiabatic case where the system is well described by a particle
characterized with a weakly renormalized mass but a small spectral
weight. In the same regime, further increasing the $el-ph$
coupling, the renormalized mass shows a smooth increase, while the
average kinetic energy is not strongly reduced. Finally we have
proposed a "phase diagram" according to the values assumed by the
spectral weight. It is found that there is strong mixing between
electronic and phononic degrees of freedom for small values of the
$el-ph$ coupling constant in the adiabatic case.

The variational approach can be easily generalized to high
dimensions, \cite{17} and it has been recently applied to study
the three dimensional continuum Fr${\ddot o}$hlich model giving a
very good description of ground state features. \cite{giulio} In
any case the results discussed in this paper are not limited to
the one-dimensional case as confirmed by the behavior of some
properties on the square lattice. \cite{alex,alex1}

\section*{Figure captions}
\begin {description}

\item{Fig.1}
The ground state energy $E_0$ in units of $\omega_0$ (a), the
spectral weight Z (b), the average kinetic energy $K$ (c) and the
average phonon number $N$ (d) as a function of the coupling
constant $\alpha$ for different values of the adiabatic ratio:
$\omega_0 /t =2$ (solid line), $\omega_0 /t=1$ (dashed line),
$\omega_0 /t =0.5$ (dotted line) and $\omega_0 /t=0.25$
(dash-dotted line). The diamonds in figure (a) indicate the $PIMC$
data for the energy kindly provided by P. E. Kornilovitch at
$\omega_0 /t =1$, and the squares on a dash-dotted line in figure
(b) denote the ratio $m/m^*$ obtained within the variational
approach at $\omega_0 /t =0.25$.

\item{Fig.2}
(a) The electron-lattice  correlation function $S(R_l)$ at
$\omega_0 /t =1$ for different values of the coupling:
$\lambda=0.5$ (circles), $\lambda=1.25$ (squares), $\lambda=2.0$
(diamonds), $\lambda=2.75$ (triangles up), and $\lambda=3.5$
(triangles down). In the inset the normalized correlation function
$\chi(R_l)$ at $\omega_0 /t =1$ for $\lambda=0.5$ (circles) and
$\lambda=2.75$ (squares).

(b) The electron-lattice correlation function $S(R_l)$ at $\alpha
=2$ for different values of the adiabatic parameter: $\omega_0 /t
=2$ (circles), $\omega_0 /t =1$ (squares), $\omega_0 /t =0.5$
(diamonds), and $\omega_0 /t =0.25$ (triangles up).

\item{Fig.3}
Polaron "phase diagram" for long-range (solid line) and Holstein
(dashed line) $el-ph$ interaction. The transition lines correspond
in weak coupling to model parameters such that the spectral weight
$Z=0.9$, in strong coupling such that $Z=0.1$.

\item{Fig.4}
(a) The average kinetic energy $K$ as a function of the constant
coupling $\lambda$ at $\omega_0 /t =1$ for  long-range (solid
line) and local Holstein $el-ph$ interaction (dashed line).

(b) The average kinetic energy $K$ as a function of the adiabaticity ratio
$\omega_0 /t$ for  long-range (solid line) and
Holstein (dashed line) $el-ph$ interaction in correspondence of model parameters
such that the spectral weight $Z=0.1$.

\end {description}

\end {document}
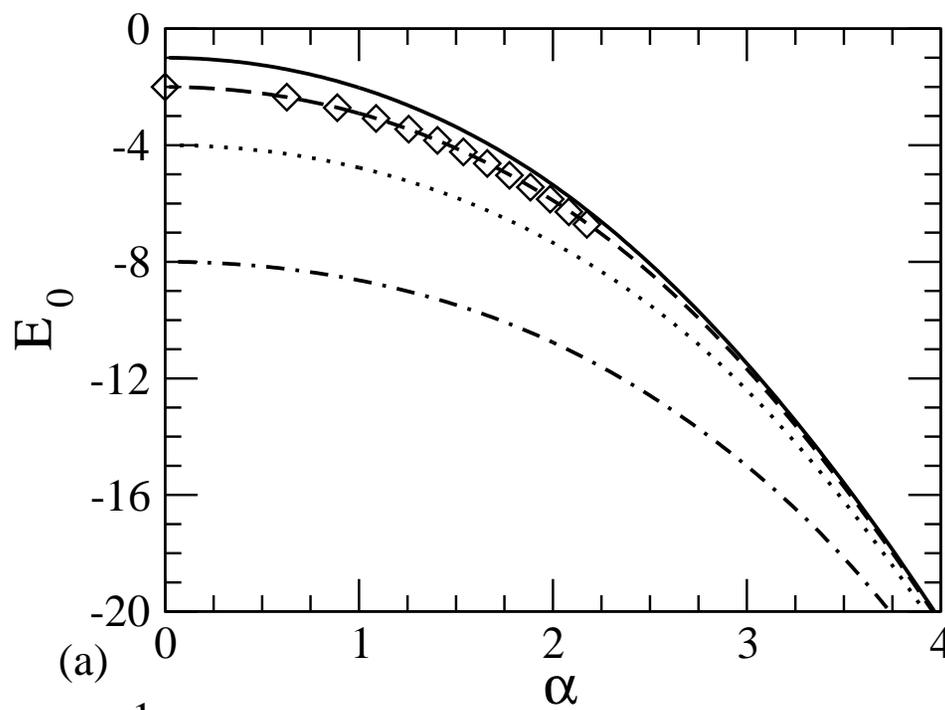
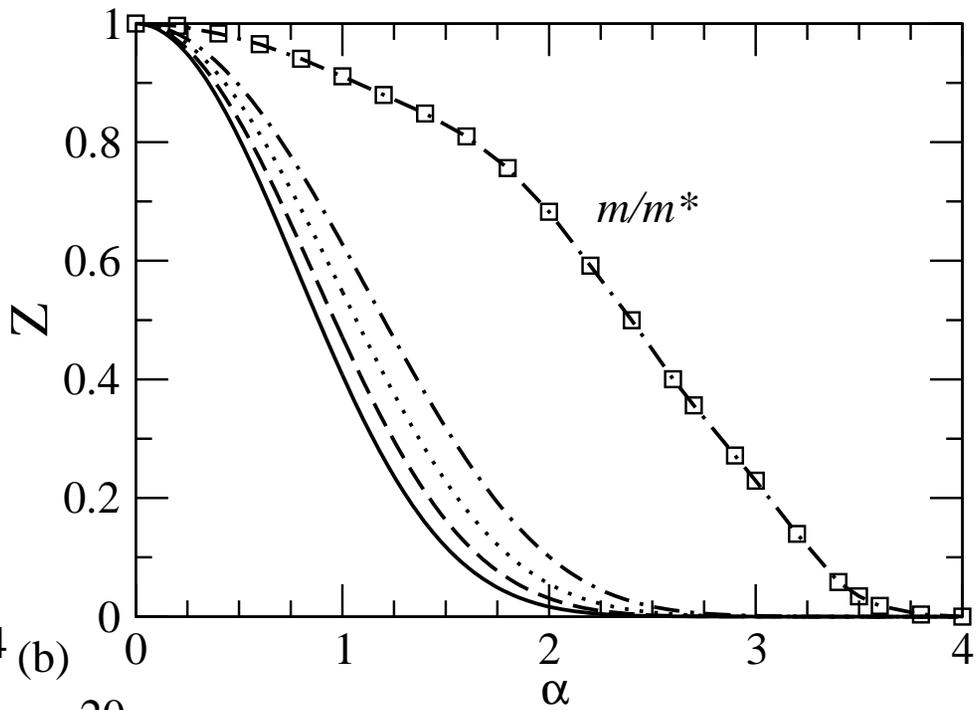
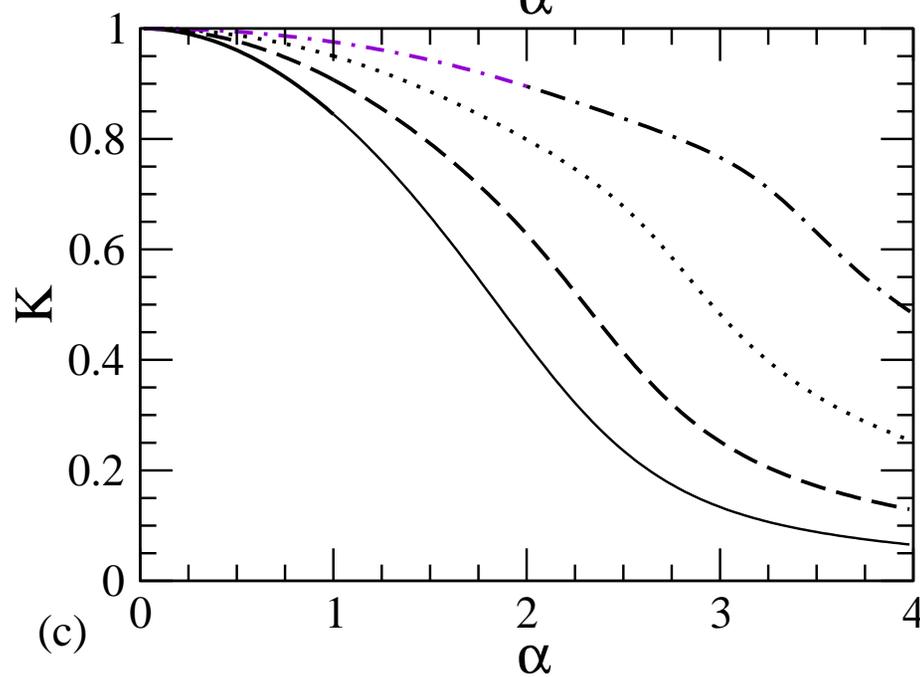
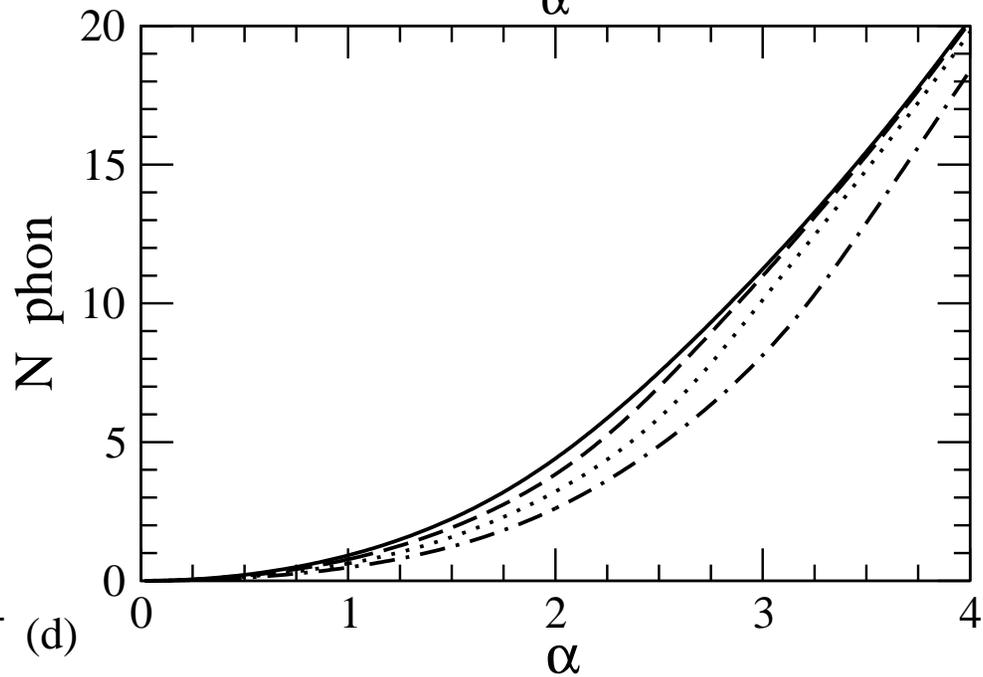
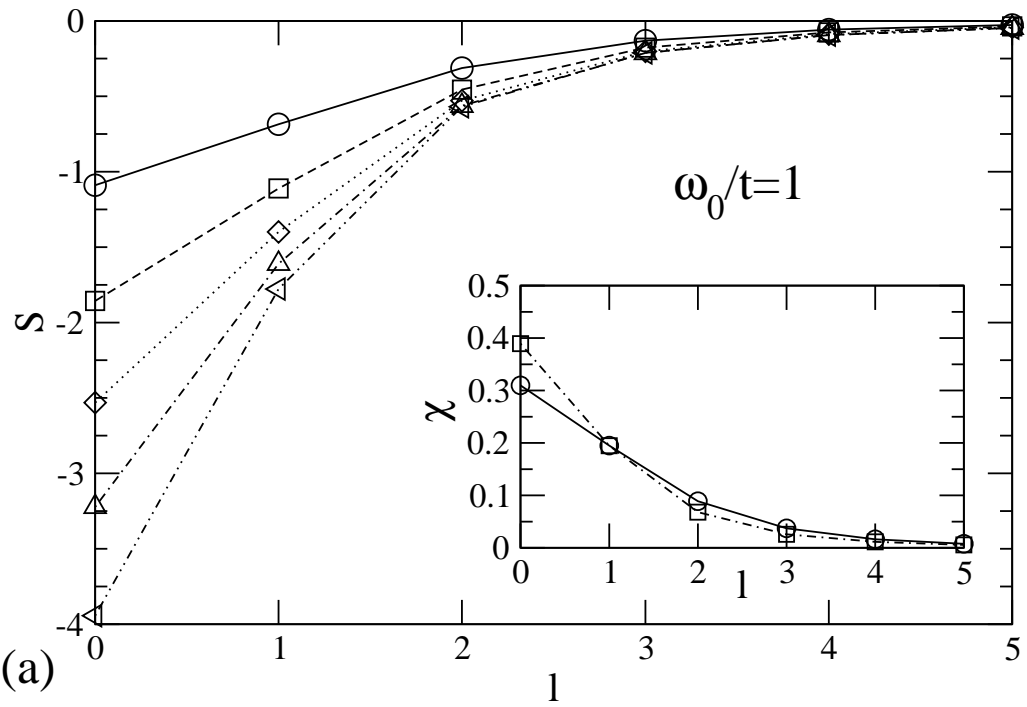
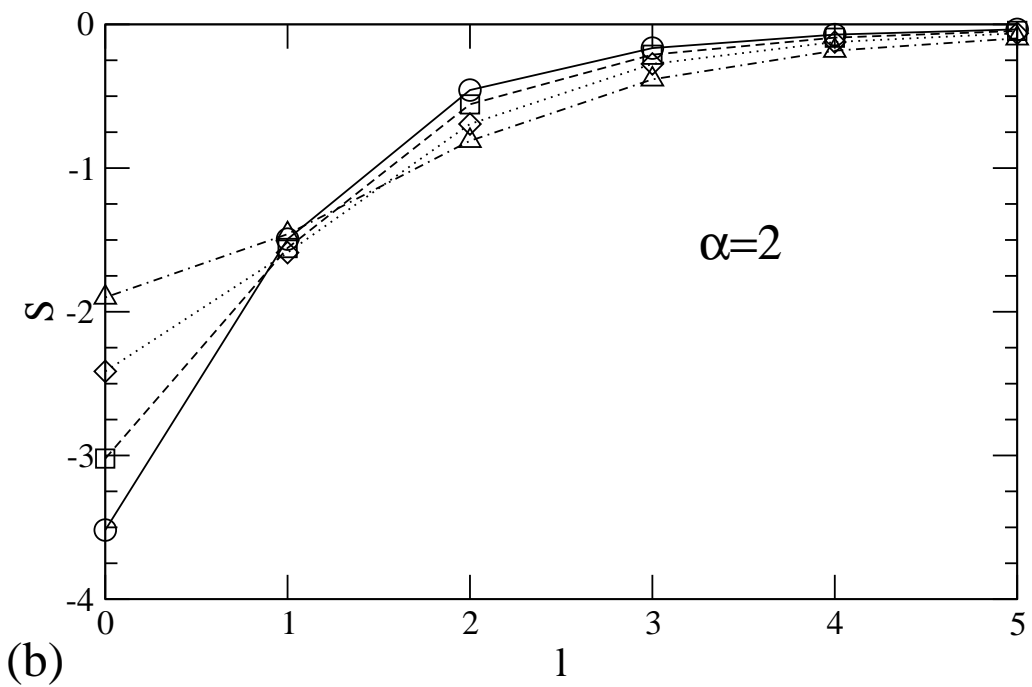
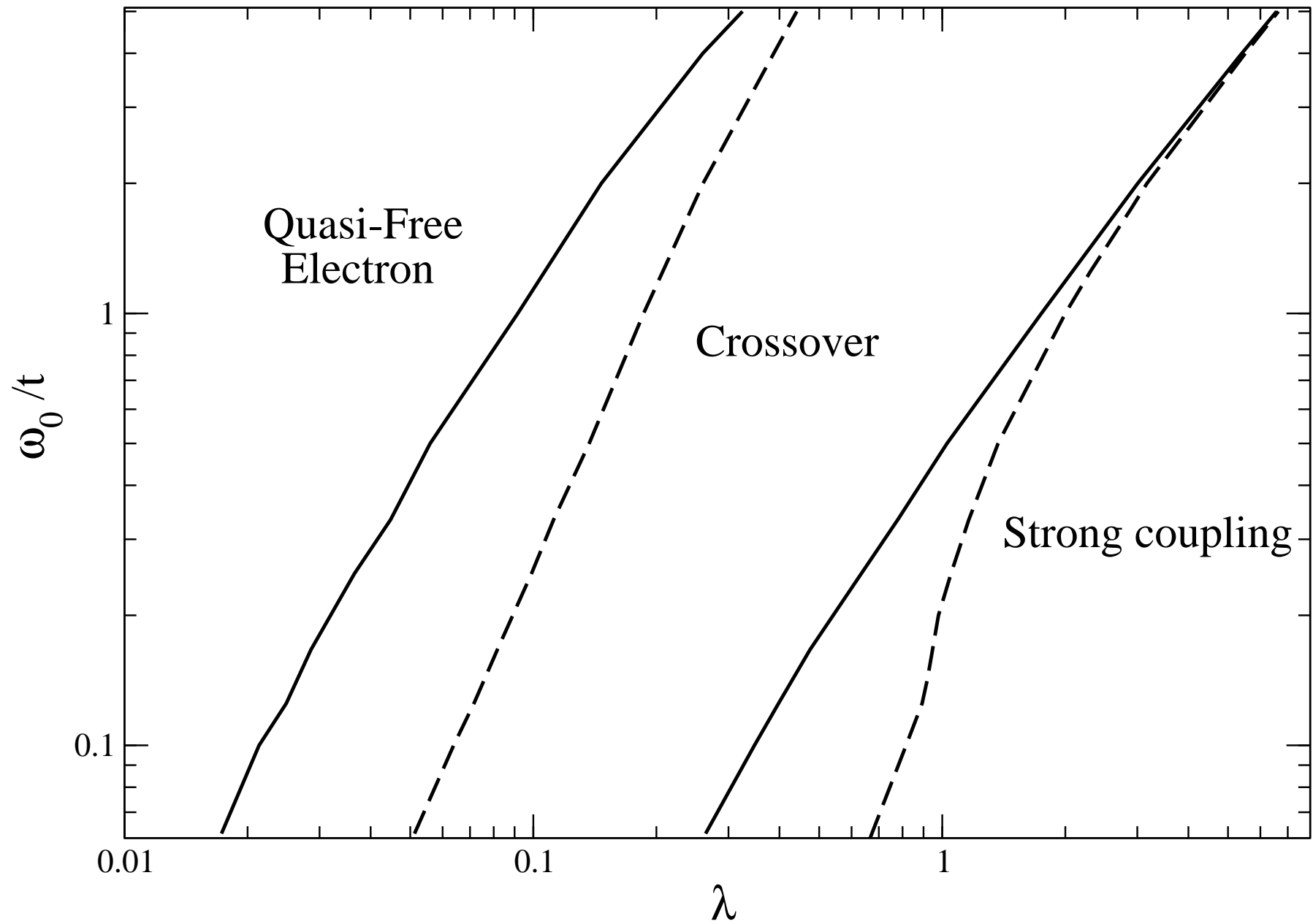
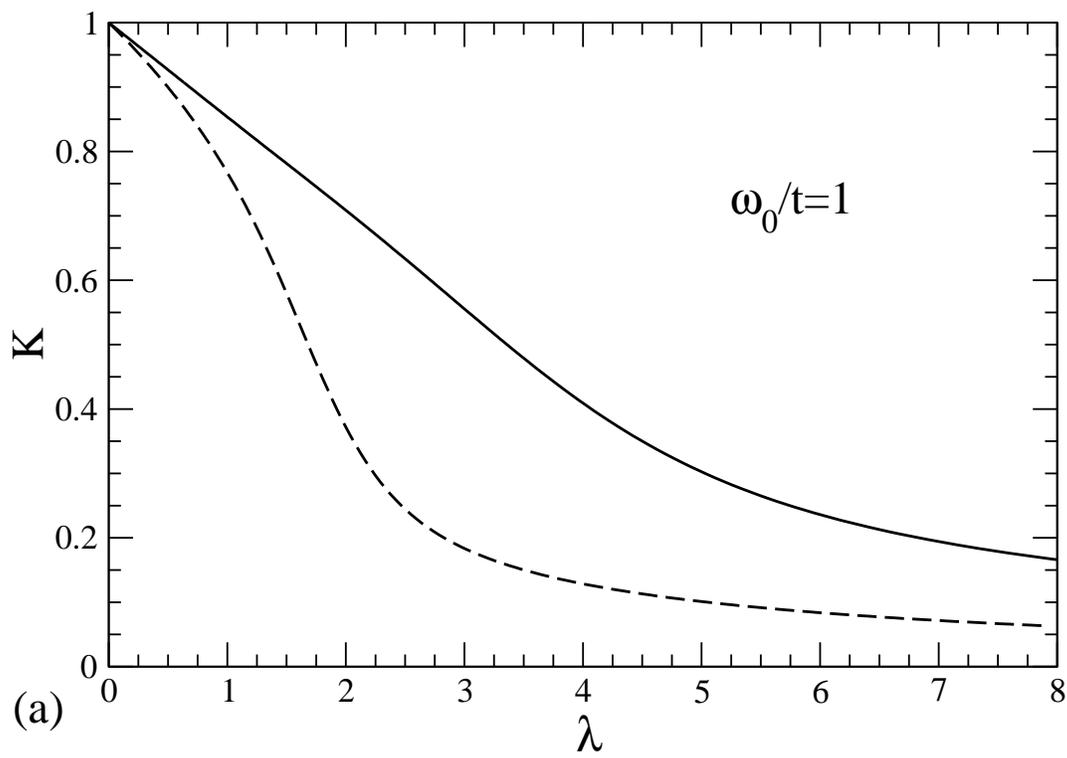
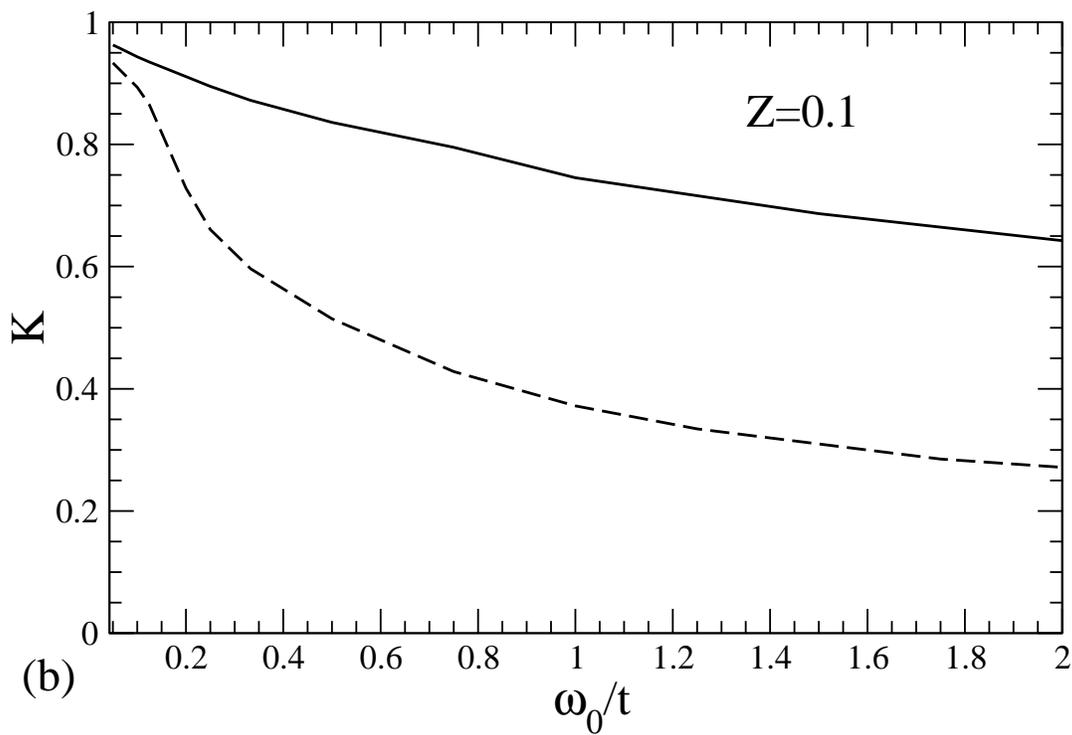
